\documentclass[pre, twocolumn, showpacs, superscriptaddress]{revtex4-1}
\usepackage{amsmath, amssymb}
\usepackage{graphicx}
\begin{document}

\title{Inhomogeneity of the phase space of the damped harmonic oscillator under L\'evy noise}
\author{Zhan Cao}
\affiliation{Center for Interdisciplinary Studies $\&$ Key Laboratory for
Magnetism and Magnetic Materials of the MoE, Lanzhou University, Lanzhou 730000, China}
\author{Yu-Feng Wang}
\affiliation{Center for Interdisciplinary Studies $\&$ Key Laboratory for
Magnetism and Magnetic Materials of the MoE, Lanzhou University, Lanzhou 730000, China}
\author{Hong-Gang Luo}
\affiliation{Center for Interdisciplinary Studies $\&$ Key Laboratory for
Magnetism and Magnetic Materials of the MoE, Lanzhou University, Lanzhou 730000, China}
\affiliation{Beijing Computational Science Research Center, Beijing 100084, China}

\begin{abstract}
The damped harmonic oscillator under symmetric L\'{e}vy white noise shows inhomogeneous phase space, which is in contrast to the homogeneous one of the same oscillator under the Gaussian white noise, as shown in a recent paper [I. M. Sokolov, W. Ebeling, and B. Dybiec, Phys. Rev. E \textbf{83}, 041118 (2011)]. The inhomogeneity of the phase space shows certain correlation between the coordinate and the velocity of the damped oscillator under symmetric L\'{e}vy white noise. In the present work we further explore the physical origin of these distinguished features and find that it is due to the combination of the damped effect and heavy tail of the noise. We demonstrate directly this in the reduced coordinate $\tilde{x}$ versus velocity $\tilde{v}$ plots and identify the physics of the anti-association of the coordinate and velocity.
\end{abstract}


\maketitle

\section{Introduction} \label{intr}
The harmonic oscillator is one of several basic models that can be exactly solved in textbooks and everyone in the physical community is familiar with it. The harmonic oscillator can be taken as the most basic model to describe the perturbation of an interacting system around its equilibrium positions. In reality, the harmonic oscillators are always under certain environments and their influences can be taken into account as the damping of the oscillator. However, besides the stable influence of the environment, there exists some random perturbations around the system, which are usually considered as the Gaussian white noises, and the dynamics of the system is thus physically stochastic. The characteristics of the Gaussian white noises are that the noises occur in a finite time domain (``white") and the strengths of the noises follow a Gaussian distribution. The physics of the Gaussian white noises has been intensively studied in the past decades. However, in nature many noises do not follow the Gaussian distribution but show heavy-tailed distributions, for example, a $\alpha$-stable L\'evy distribution \cite{West1987,Viecelli1993, Shlesinger1994,Janicki1994,Zimbardo1995}. Physically, the L\'evy noise results from the strongly collisions between the test particles and/or the nonequilibrated heat reservoir abruptly disturbed with a local heating to a very high temperature \cite{Baranger2002, Dybiec2004}. Mathematically, in the L\'evy noise the distribution of the random variables has a diverging variance. In comparison to the conventional Gaussian noises, the heavy-tailed noises lead to many novel phenomena observed in some realistic systems \cite{Dybiec2007,Dybiec2010a,Dybiec2010b,Chechkin2001,Chechkin2003,Chechkin2007,Chechkin2008,Ditlevsen1999a,Ditlevsen1999b,Yi-Zhong Zhuo2005,Patel1993,Olemskoi2010}.

In a recent paper \cite{Sokolov2011}, Sokolov \textit{et al.} investigated the phase space property of a damped(including underdamped and overdamped cases) harmonic oscillator under a symmetric L\'evy white noise. They found that in contrast to the homogeneity of the phase space under Gaussian white noise, the phase space of the damped harmonic oscillator under the general symmetric L\'evy white noises is inhomogeneous, and the inhomogeneous extent of the phase space is dependent of a parameter $\alpha$, which describe how far the L\'evy distribution is away from the Gaussian one. The inhomogeneity of the phase space shows physically certain dependence between the coordinate and the velocity, which is lack in the conventional Gaussian white noise case. Actually, this problem was first addressed and solved in \cite{West1982}. Their investigation showed that the variance of the velocity $v(t)$ of the damped oscillator is infinite and imply the particle has infinite kinetic energy. As a result, the fluctuation-dissipation relation, which can connect the fluctuations about equilibrium with the susceptibility of the system to external perturbation, is violated in this system. It is the breakdown of the fluctuation-dissipation relation in systems with heavy-tailed distributions that make studying the details of the L\'evy noise oscillator worth investigating in the community. In the present work we revisit this system and further explore the inhomogeneity of the phase space. We find that the inhomogeneity can be attributed to the interplay between the damped oscillating and the L\'evy noise. To confirm it, we show directly the sample points $(\tilde{x},\tilde{v})$ obtained numerically in the phase space and observe the dependence of the $\tilde{x}-\tilde{v}$. Here the $\tilde x$ and $\tilde v$ are reduced coordinate and velocity defined by $\tilde x = x/\sigma_x$ and $\tilde v = v/\sigma_v$, where $\sigma_x$ and $\sigma_v$ are the widths of the distributions of the coordinate and velocity \cite{Sokolov2011,Chandrasekhar1943}, which are also defined below. The result gives a physically reasonable interpretation of the inhomogeneity obtained in the previous publication \cite{Sokolov2011}.

 The paper is organized as follows. In section \ref{sec2} we present the model and its time-evolution dynamics. In section \ref{sec3} we discuss the inhomogeneity observed in the phase space according time-evolution dynamics. Section \ref{sec4} is devoted to a brief summary.

\begin{figure}[ht]
\begin{center}
\includegraphics[width=0.8\columnwidth]{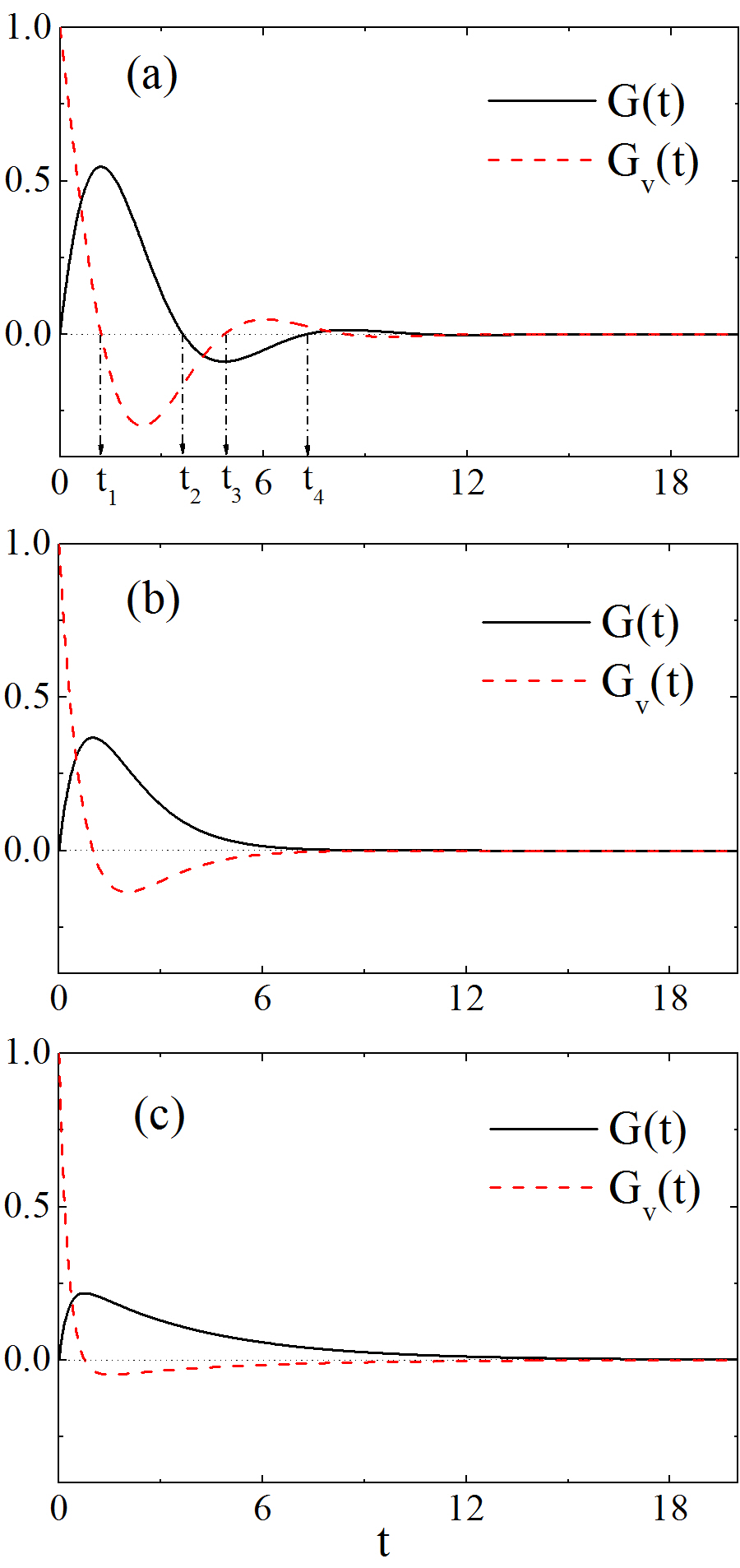}
\caption {Time-evolution of the Green's functions $G(t)$ and $G_v(t)$ for the damped harmonic oscillator with different damped cases. (a) The underdamped case ($\omega=1, \gamma=1$), (b) the critical case ($\omega=1, \gamma=2$), and (c) the overdamped case ($\omega=1, \gamma=4$). The $t_k (k = 1, 2, \cdots) $ in (a) denotes the time at which the product of $G(t)*G_v(t)$ changes sign.}\label{fig1}
\end{center}
\end{figure}

\section{Model and its time-evolution}\label{sec2}
The problem we are interested in is the phase space property of a damped harmonic oscillator under a L\'evy white noise. Here and hereafter, the L\'evy white noise is limited to the symmetric case, as discussed in Ref. \cite{Sokolov2011}. The model can be described by the Langevin equation of the coordinate $x(t)$
\begin{equation}
\label{eq4}
\frac{d^2}{dt^2} x(t) + \gamma \frac{d}{dt} x(t) + k x(t) = \xi(t),
\end{equation}
where $\xi(t)$ denotes the white noise satisfying the symmetric L\'evy distribution, $k$ is the elastic constant and $\gamma$ is the damped coefficient. For comparison, we use the same notations as those in Ref. \cite{Sokolov2011}. It is well-known that this model can describe some basic phenomena in the nonequilibrium statistical physics. Formally, this equation can be solved exactly as follows
\begin{eqnarray}
&& x(t) = F(t)+\int^t_0G(t-t')\xi(t')dt', \label{eq5}\\
&& v(t) = F_v(t)+\int^t_0G_v(t-t')\xi(t')dt', \label{eq6}
\end{eqnarray}
where $v(t) = \frac{d}{dt} x(t)$ denotes the velocity of the oscillator and $F(t)$ and $F_v(t)$ are decaying functions dependent of the initial conditions, respectively. Here $G(t)$ and $G_v(t)$ are the Green's (response) functions governing the time evolution of the system, and they have the relation $ G_v(t)= \frac{d}{dt}G(t) $. It is convenient to take $t-t' \rightarrow t'$, Eqs. (\ref{eq5}) and (\ref{eq6}) can be rewritten as
\begin{eqnarray}
&& x(t)=F(t)+\int^t_0G(t')\xi(t-t^{\prime})dt^{\prime } , \label{eq7}\\
&& v(t)=F_v(t)+\int^t_0G_v(t')\xi(t-t^{\prime})dt^{\prime }. \label{eq8}
\end{eqnarray}
By using the Laplace transformation, one can obtain analytically the expressions of $G(t)$ and $G_v(t)$, which are given at different cases.

Case I: The underdamped case, one has $\omega=\sqrt{\kappa}>\gamma/2$. Defining $\Omega=\sqrt{\omega^2-\gamma^2/4}$, $G(t)$ and $G_v(t)$ read, respectively,
\begin{eqnarray}
&& G(t) =\frac{\text{exp}(-\gamma t/2)}{\Omega}\sin(\Omega t), \label{9}\\
&& G_v(t) =\text{exp}(-\gamma t/2)\left[\cos(\Omega t)-\frac{\gamma \sin(\Omega t)}{2\Omega}\right]. \label{eq10}
\end{eqnarray}
The typical behaviors of the time evolutions of the $G(t)$ and $G_v(t)$ are decaying oscillatingly, as shown in Fig. \ref{fig1}(a).

Case II: The critical case, one has $\omega=\sqrt{\kappa}=\gamma/2$, $G(t)$ and $G_v(t)$ read, respectively,
\begin{eqnarray}
&& G(t) = t \text{exp}(-\gamma t/2), \label{eq11}\\
&& G_v(t) = \text{exp}\left(-\gamma t/2\right)(1-\gamma t/2). \label{eq12}
\end{eqnarray}

Case III: The overdamped case, one has $\omega=\sqrt{\kappa}<\gamma/2$. Defining $\Omega=\sqrt{\gamma^2/4-\omega^2}$, $G(t)$ and $G_v(t)$ read, respectively,
\begin{eqnarray}
&& G(t) = \frac{\text{exp}(-\gamma t/2)}{\Omega}\sinh(\Omega t), \label{eq13}\\
&& G_v(t) = \text{exp}(-\gamma t/2)\left[\cosh(\Omega t)-\frac{\gamma \sinh(\Omega t)}{2\Omega}\right]. \label{eq14}
\end{eqnarray}
In both the critical and the overdamped cases, $G(t)$ and $G_v(t)$ decay without oscillating. These time-evolution behaviors of the Green's functions under different cases play an important role in understanding the inhomogeneity of the phase space of the damped harmonic oscillator under the L\'evy noise, as discussed later.

\section{Phase space properties of the damped oscillator under L\'evy noise} \label{sec3}
Following Ref. \cite{Sokolov2011}, the characteristic function of the joint probability density in the phase space of the damped harmonic oscillator at time $t$ is expressed as
\begin{equation} \label{eq15}
f(k,q,t) = \text{exp}\left[-\sigma^\alpha \int^t_0 |kG(t')+qG_v(t')|^\alpha dt'\right]
\end{equation}
where $\alpha$ is the L\'{e}vy index, $\sigma$ is the scale parameter. The corresponding joint probability density $p(x,v,t)$ is
\begin{equation} \label{eq16}
p(x,v,t) = \frac{1}{4\pi^2} \int^\infty_{-\infty} dk \int^\infty_{-\infty} dq f(k,q,t)\text{exp}[-i(kx+qv)]
\end{equation}

Due to the damping, in long enough time $G(t)$ and $G_v(t)$ approach to zero for all three cases, as shown in Fig.\ref{fig1} and thus the value of $f(k,q,t)$ for given $k$ and $q$ reaches a stable value. In this case, it means that the joint probability density $p(x,v,t)$ also reaches a fixed point in the phase space.

\begin{figure}[ht]
\begin{center}
\includegraphics[width=\columnwidth]{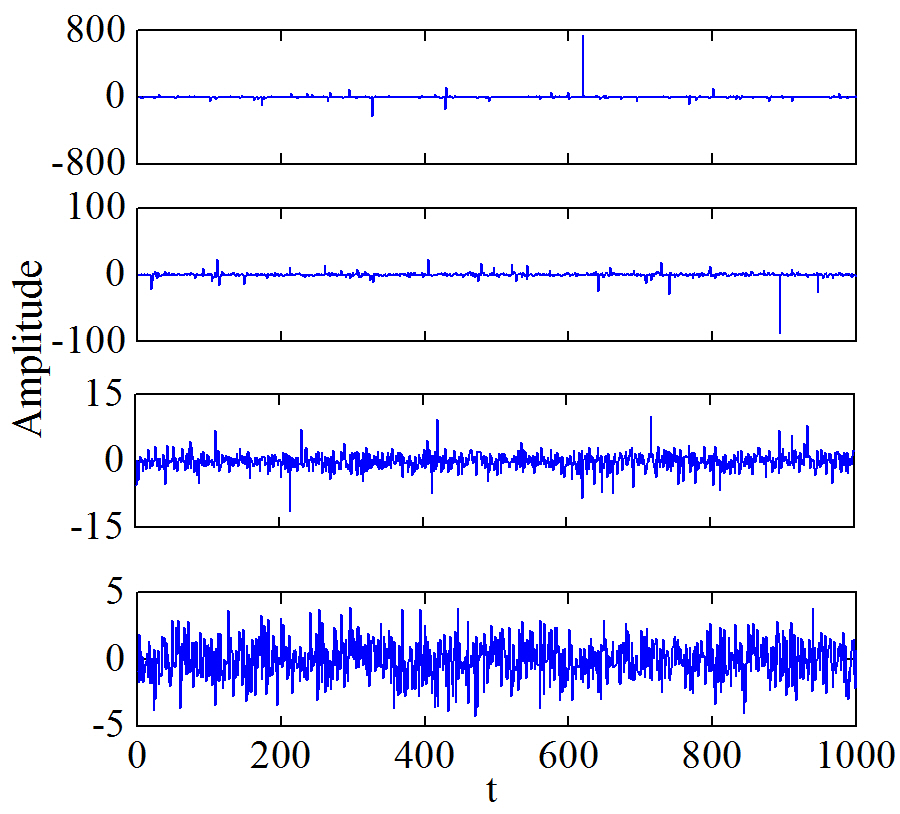}
\caption {Symmetric L\'{e}vy white noise with different indices $\alpha$ (from top to bottom, $\alpha=1.0, 1.5, 1.8, 2.0$ ).}\label{fig2}
\end{center}
\end{figure}

In the following we solve numerically Eqs. (\ref{eq7}) and (\ref{eq8}). In the process we consider a L\'{e}vy noise $\xi_{\alpha,D}(t)$ defining by $\int^{t+\Delta t}_t\xi_{\alpha,D}(t^\prime)dt^\prime \equiv L_{\alpha,D}(\Delta t)$, where the value of $L_{\alpha,D}(\Delta t)$ is given in a L\'{e}vy process with characteristic function as follows
\begin{eqnarray}
&& f_{\alpha,D}(k,\Delta t) = \int^\infty_{-\infty} \textrm{exp}(i k L_{\alpha,D})p(L_{\alpha,D},\Delta t)dL_{\alpha,D}\nonumber\\
&& \hspace{1.7cm}= <\textrm{exp}(ikL_{\alpha,D})>=\textrm{exp}(-D|k|^\alpha \Delta t). \label{eq18}
\end{eqnarray}
Here $D = \sigma^\alpha$ denotes the strength of noise. It is well-known that $L_{\alpha,D}$ has a scale behavior, namely, $L_{\alpha,D}=D^{1/\alpha}L_{\alpha,1}$ \cite{Sliusarenko2007}. In Fig. \ref{fig2} the explicit behaviors of the L\'evy noise are shown for different indices. For the method of generating L\'{e}vy noise one can refer to Ref. \cite{Chambers1976,Leccardi2000}.

\begin{figure}[ht]
\begin{center}
\includegraphics[width=\columnwidth]{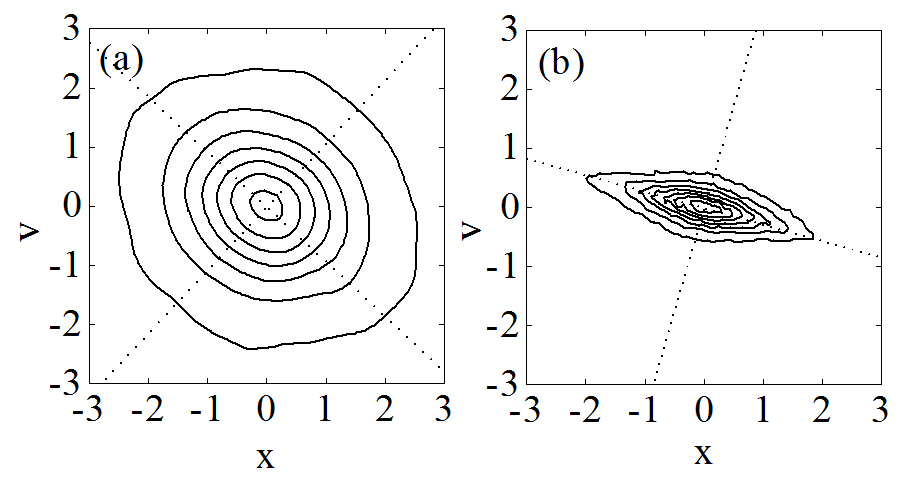}
\caption {The joint probability density $p(x,v)$ for $\alpha=1$. The results are obtained statistically from $200 000$ sample points. (a) The underdamped case ($\omega=1, \gamma=1$) and (b) the overdamped case ($\omega=1, \gamma=4$). }\label{fig3}
\end{center}
\end{figure}

\begin{figure}[ht]
\begin{center}
\includegraphics[width=\columnwidth]{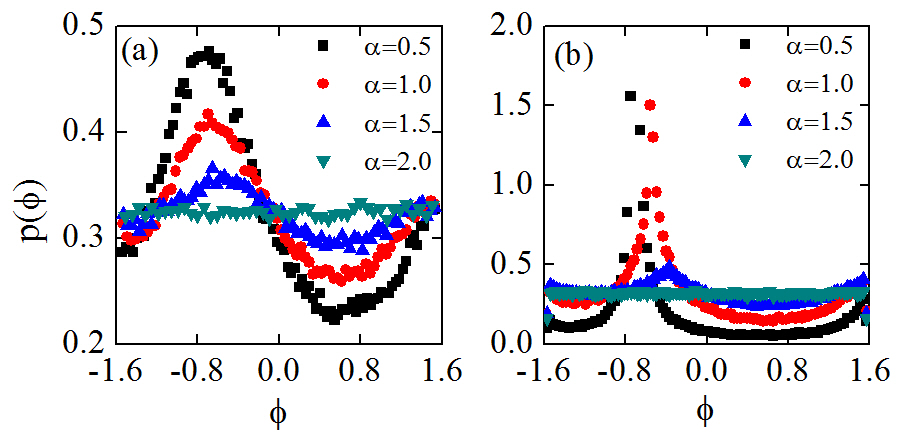}
\caption {Simulated phase distribution $p(\phi)$ for different values of $\alpha$. (a) The underdamped case ($\omega = 1,\gamma = 1$) and (b) the overdamped case ($\omega = 1,\gamma = 4$).}\label{fig4}
\end{center}
\end{figure}

Equations (\ref{eq7}) and (\ref{eq8}) are integrated numerically according to the following discrete version
\begin{eqnarray}
&& x(N\Delta t) = F(N\Delta t)+(D\Delta t)^{\frac{1}{\alpha}}\sum^{N-1}_{i=0}G(i\Delta t)\zeta_{\alpha,1}(N-i), \label{eq19}\\
&& v(N\Delta t) = F_v(N\Delta t)+(D \Delta t)^{\frac{1}{\alpha}}\sum^{N-1}_{i=0}G_v(i\Delta t)\zeta_{\alpha,1}(N-i), \label{eq20}
\end{eqnarray}
where $\zeta_{\alpha,1}(i)$ is the \textit{i}th number of the random numbers array, which obeys the symmetric L\'{e}vy distribution with index parameter $\alpha$ and scale parameter unity. More details on the numerical scheme for integration of stochastic differential equations with respect to $\alpha$-stable noises can be found in \cite{Janicki1994,Sliusarenko2007,Dybiec2006,Janicki1996}. For simplicity, we just study the unity noise, i.e. $D=1$. Eqs. (\ref{eq19}) and (\ref{eq20}) are simulated by taking initial condition $x(0) = 0, v(0) = 1$, time-step $\Delta t= 0.01$ and total $N = 2000$ steps to guarantee time-evolution convergence. We do repeatedly $200 000$ times time-evolution of $x(t)$ and $v(t)$ and thus obtain $200 000$ sample points in the phase space, which have been analyzed statistically. To confirm our numerical approach, we reproduce the stationary joint probability density $p(x,v)$ for $\alpha = 1$, as shown in Fig. \ref{fig3} for the underdamped and the overdamped cases. One can note that these results are in good agreement with those in Ref. \cite{Sokolov2011}. We also obtained the phase distribution $p(\phi)$ in Fig. \ref{fig4} for different L\'evy indices, which is also consistent with that presented in Ref. \cite{Sokolov2011}. The angle $\phi$ here is defined as $\phi=\arctan(\tilde{v}/\tilde{x})$.

In Ref. \cite{Sokolov2011}, the inhomogeneity of the joint probability density was attributed to certain association of the coordinate and velocity of the harmonic oscillator under a L\'evy noise. Some characteristic features are observed. They are that i) the distribution of the coordinate and velocity is not elliptic, which is in contrast to that under the Gaussian white noise; ii) the phase angle shows a nontrivial distribution; and iii) the dependence of the reduced coordinate $\tilde{x} = x/\sigma_x$ and the reduced velocity $\tilde{v} = v/\sigma_v$ is stronger in the overdamped case than in the underdamped one. Here $\sigma_x=[\int^\infty_0 |G(t)|^\alpha dt]^{1/\alpha}$, $\sigma_v=[\int^\infty_0 |G_v(t)|^\alpha dt]^{1/\alpha}$. Likewise, the phase angle shows more exotic behavior in the overdamped case than in the underdamped case. These results were completely attributed to the presence of L\'evy noise. In the following we explore carefully the property of the phase space and identify the essential physics of these nontrivial phase space distributions.
\begin{figure}[ht]
\begin{center}
\includegraphics[width=\columnwidth]{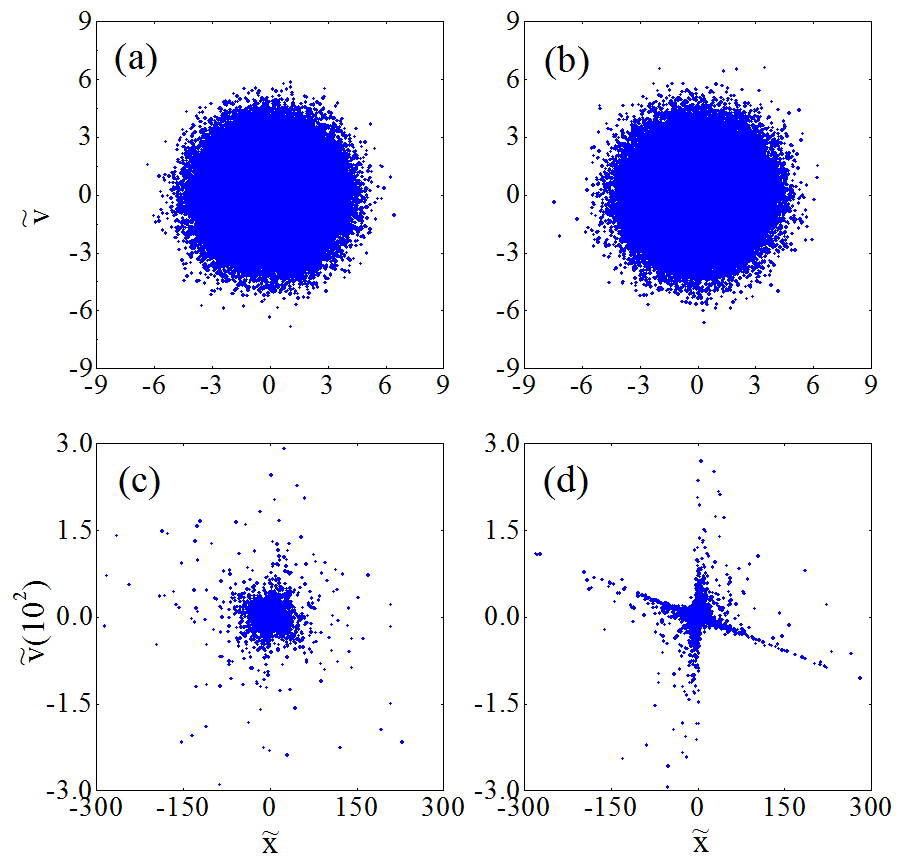}
\caption {Direct distribution of $200 000$ sample points in the phase space for different cases. (a) The underdamped case ($\omega=1, \gamma=1$) and (b) the overdamped case ($\omega=1, \gamma=4$) with Gaussian white noise ($\alpha = 2$); (c) the underdamped case ($\omega=1, \gamma=1$) and (d) the overdamped case ($\omega=1, \gamma=4$) with L\'evy noise ($\alpha = 1.5$). The initial condition is taken as $x(0) = 0, v(0) = 1$, the time-step $\Delta t=0.01$ and the time-evolution steps $N = 2000$. }\label{fig5}
\end{center}
\end{figure}
\begin{figure}[ht]
\begin{center}
\includegraphics[width=\columnwidth]{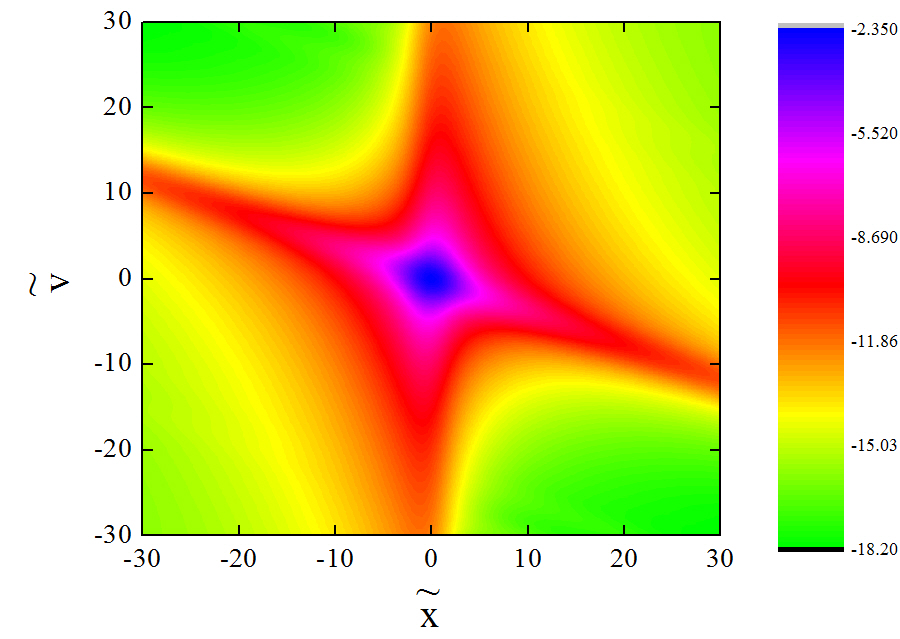}
\caption {Contour of the joint probability density $p(\tilde{x},\tilde{v})$ with logrithm scale for Fig. \ref{fig5}(d) with the same parameters.}\label{fig6}
\end{center}
\end{figure}
\begin{figure}[ht]
\begin{center}
\includegraphics[width=\columnwidth]{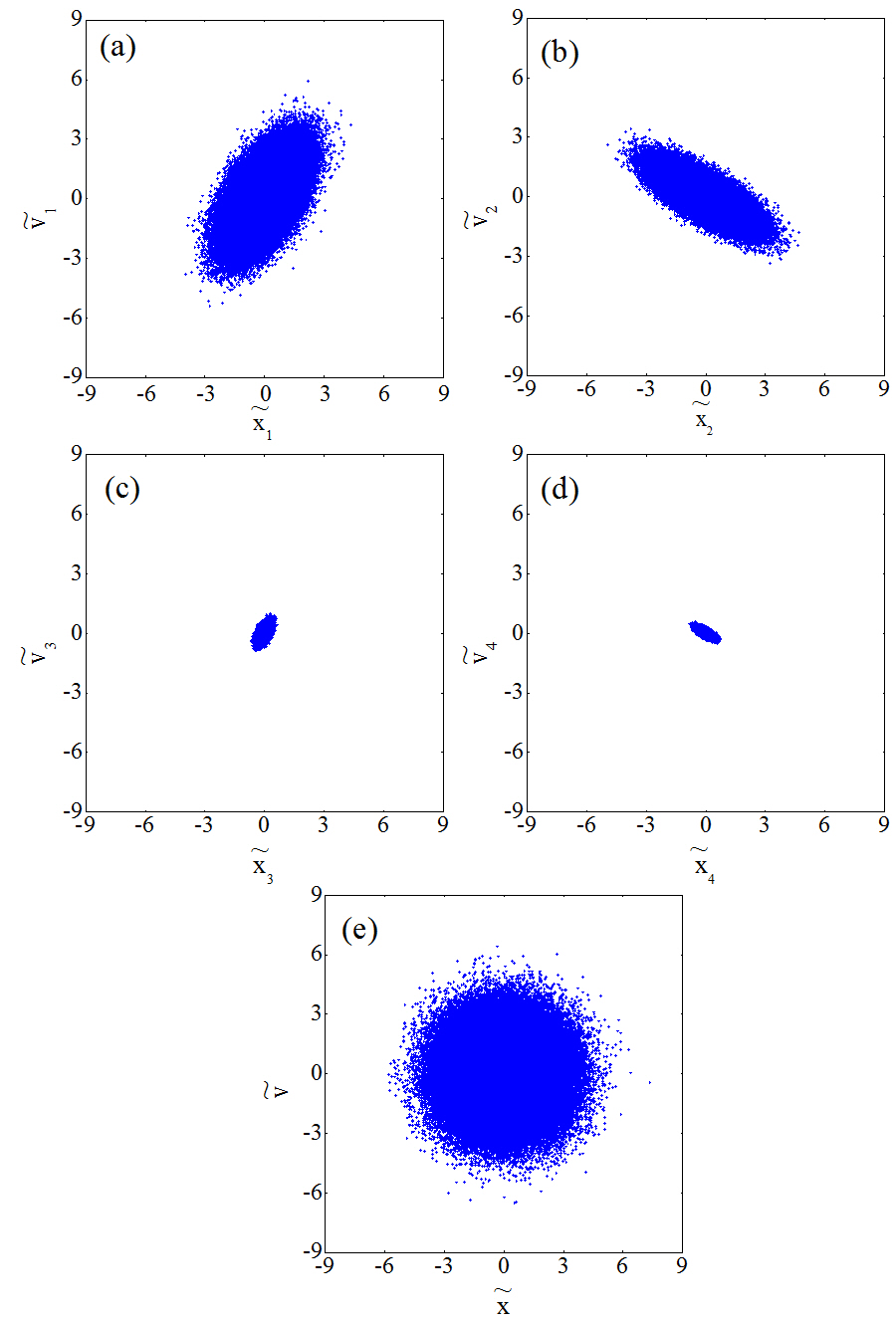}
\caption {(a)-(d) Distribution of the sample points $(\tilde{x}_k, \tilde{v}_k)(k = 1, \cdots, 4)$ at different time intervals as explained in the context and (e) the total distribution of $(\tilde{x},\tilde{v})$ after a long enough time-evolution for the underdamped oscillator ($\omega=1, \gamma=1$) under the Gaussian white noise. The initial condition is taken as $x(0) = 0, v(0) = 1$, the time-step $\Delta t=0.01$ and the time-evolution steps $N = 2000$. The time intervals are shown in Fig. \ref{fig1}(a). }\label{fig7}
\end{center}
\end{figure}
\begin{figure}[h]
\begin{center}
\includegraphics[width=\columnwidth]{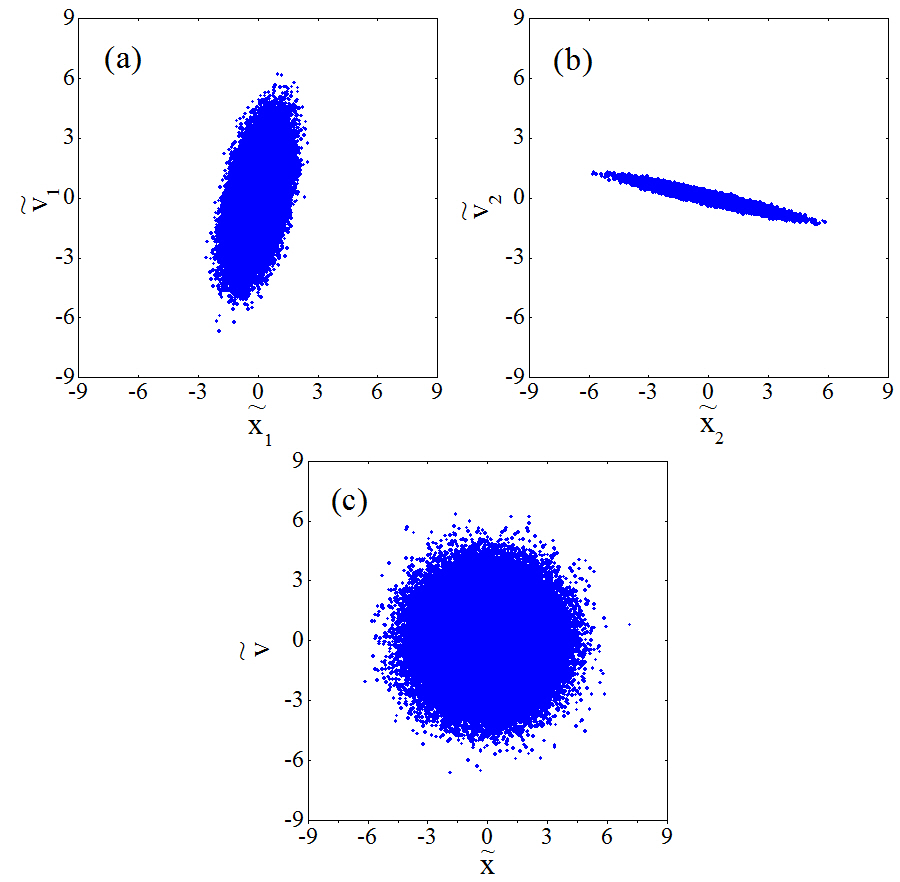}
\caption {(a) and (b) Distribution of the sample points $(\tilde{x}_k, \tilde{v}_k)(k = 1, 2)$ at different time intervals and (c) the total distribution of $(\tilde{x},\tilde{v})$ after a long enough time-evolution for the overdamped oscillator ($\omega=1, \gamma=4$) under the Gaussian white noise. The initial condition is taken as $x(0) = 0, v(0) = 1$, the time-step $\Delta t=0.01$ and the time-evolution steps $N = 2000$. }\label{fig8}
\end{center}
\end{figure}
\begin{figure}[ht]
\begin{center}
\includegraphics[width=\columnwidth]{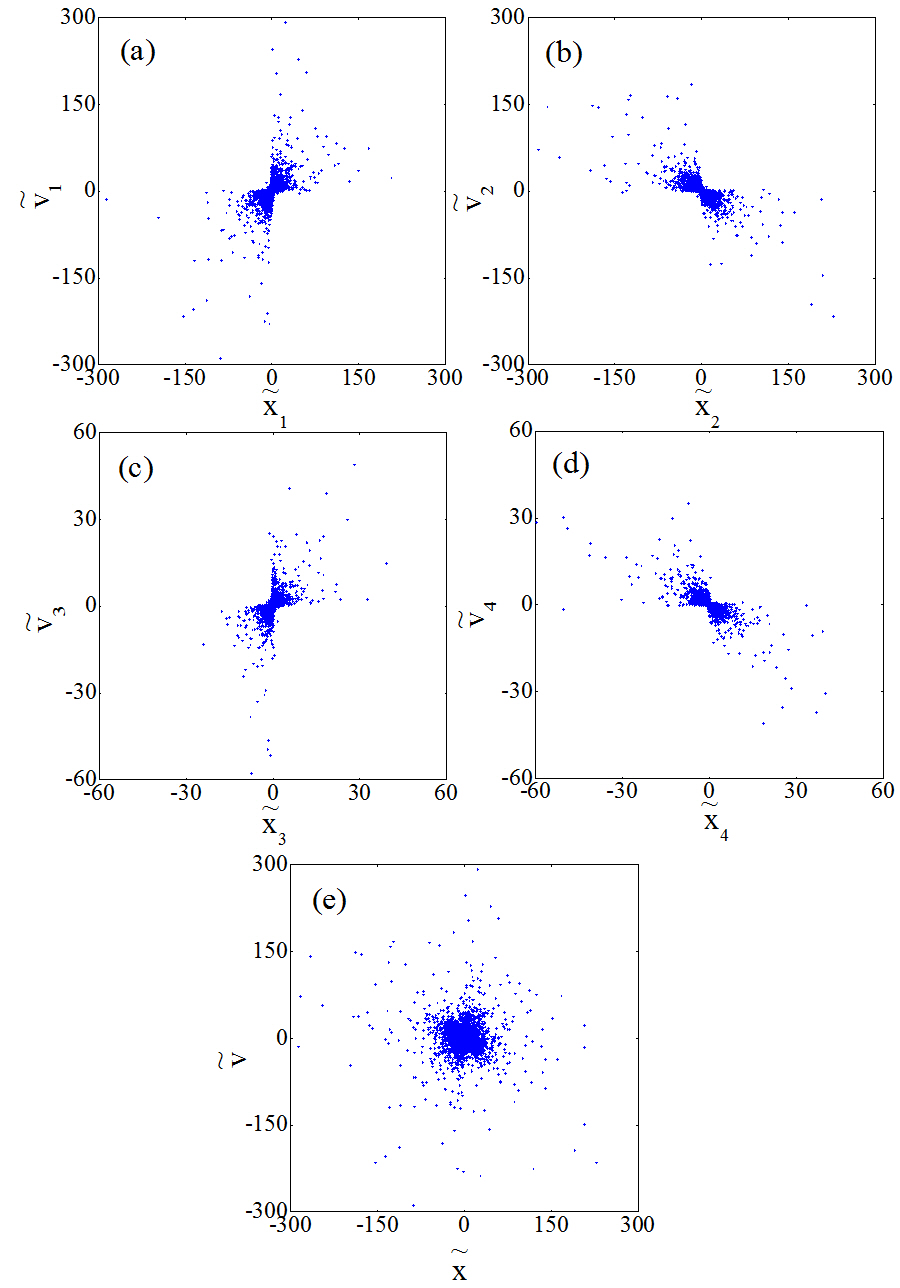}
\caption {(a)-(d) Distribution of the sample points $(\tilde{x}_k, \tilde{v}_k)(k = 1, \cdots, 4)$ at different time intervals as explained in the context and (e) the total distribution of $(\tilde{x},\tilde{v})$ after a long enough time-evolution for the underdamped oscillator ($\omega=1, \gamma=1$) under the L\'evy white noise with $\alpha = 1.5$. The initial condition is taken as $x(0) = 0, v(0) = 1$, the time-step $\Delta t=0.01$ and the time-evolution steps $N = 2000$. The time intervals are shown in Fig. \ref{fig1}(a).}\label{fig9}\end{center}\end{figure}
\begin{figure}[ht]
\begin{center}
\includegraphics[width=\columnwidth]{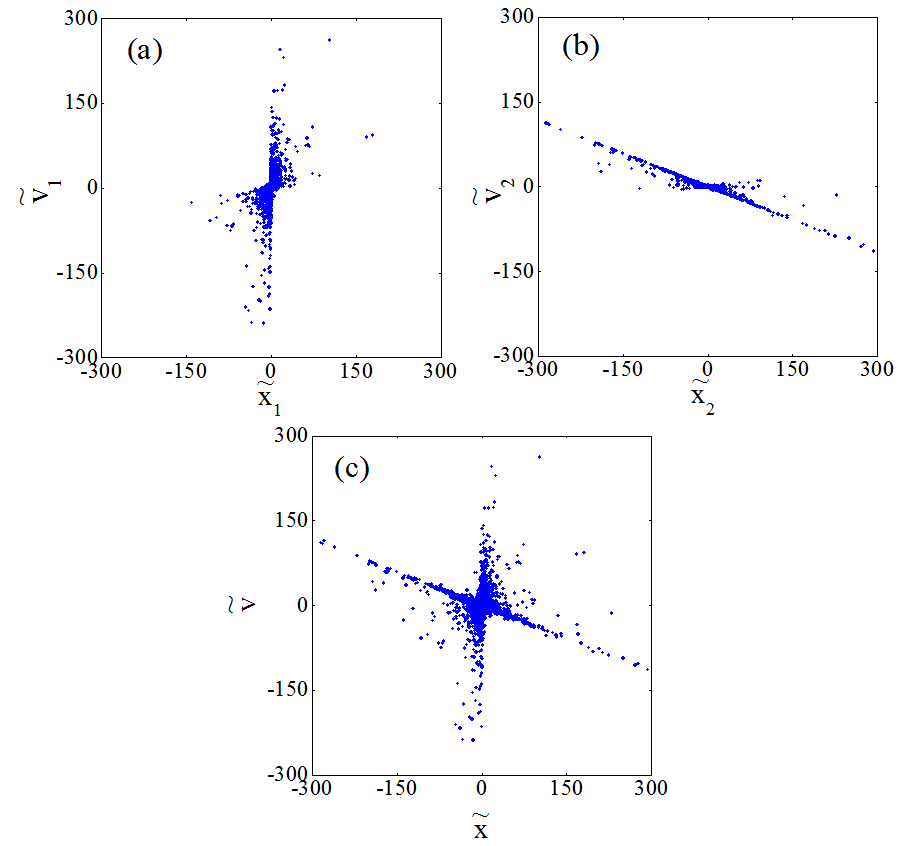}
\caption {(a) and (b) Distribution of the sample points $(\tilde{x}_k, \tilde{v}_k)(k = 1, 2)$ at different time intervals and (c) the total distribution of $(\tilde{x},\tilde{v})$ after a long enough time-evolution for the overdamped oscillator ($\omega=1, \gamma=4$) under the L\'evy white noise with $\alpha = 1.5$. The initial condition is taken as $x(0) = 0, v(0) = 1$ and the time-step $\Delta t=0.01$ and the time-evolution steps $n = 2000$.}\label{fig10}
\end{center}
\end{figure}

Without resorting to additional analysis, in Fig. \ref{fig5} we present directly $200 000$ sample points in the phase space with reduced coordinate $\tilde x$ and velocity $\tilde v$ of the damped oscillator under different white noises. We consider the underdamped case ($\omega = 1, \gamma = 1$) in Fig. \ref{fig5} (a, c) and the overdamped case ($\omega = 1, \gamma = 4$) in Fig. \ref{fig5}(b, d). For the white noises, we consider the Gaussian noise with $\alpha = 2$ and the L\'evy noise with $\alpha = 1.5$. Under the Gaussian white noise, despite of the underdamped case or the overdamped one, the distribution of the sample point in the phase space is completely homogeneous. This indicates that the random variables $\tilde{x}$ and $\tilde{v}$ are independent, which is consistent with the previous result. However, under the L\'evy noise, the distributions show quite different features. For the underdamped case, no apparent anisotropic distribution is observed but some scattered points are far away from $(\tilde{x},\tilde{v}) = (0,0)$. This result can be attributed to the heavy tail of the L\'evy noise. For the overdamped case, the distribution of the sample points in the phase space is strongly anisotropic, in particular in the regions far away from $(\tilde{x},\tilde{v})= (0, 0)$. In these regions, while the sample points can be distributed into the regions roughly satisfying $\tilde{x}(t)*\tilde{v}(t) > 0$, there is no any sample points to go into the regions with $\tilde{x}(t)*\tilde{v}(t) < 0$ in the long time limit, which is an interesting observation surprised us. This feature is also shown in Fig. \ref{fig6} near $(\tilde{x},\tilde{v})= (0, 0)$, which is numerically calculated by Eqs.(\ref{eq15}) and (\ref{eq16}). This anisotropic distribution of the sample points is just inhomogeneous phase space observed in Ref. \cite{Sokolov2011}. So, what is the physical origin of this inhomogeneity in the phase space? In the following we present a detail analysis.

We check carefully the time-evolution of the Green's function $G(t)$ and $G_v(t)$, as shown in Fig. \ref{fig1}. For clarity, we only consider the underdamped case in Fig. \ref{fig1}(a) with $\omega = 1, \gamma = 1$. When $G_v(t)$ decays with oscillating, $G(t)$ increases initially, then also decays with oscillating. So with the time-evolution, the whole time domain can be divided into many different intervals in which $G(t) * G_v(t)$ changes sign alternatively. For example, when $t \in (t_0, t_1)$ where $t_0 = 0$, $G(t) > 0$ and $G_v(t) > 0$, thus $G(t) * G_v(t) > 0$. When $ t \in (t_1, t_2) $, $G_v(t)$ changes sign, $G_v(t) < 0$ but $G(t)$ is still greater than zero, as a result,  $G(t) * G_v(t) < 0$. Furthermore, when $t \in (t_2, t_3)$,  $G(t) * G_v(t)$ is again greater than zero due to that $G(t)$ becomes negative in this time interval. Finally, when $t\in (t_3, t_4)$,  $G(t) * G_v(t)$ becomes negative since $G(t)$ and $G_v(t)$ have different signs in this time interval. Of course, the behavior of changing sign of $G(t)*G_v(t)$ can continue till $G(t)$ and $G_v(t)$ decays away. What does this observation mean? Here it is helpful to rewrite Eqs. (\ref{eq19}) and (\ref{eq20}) as
\begin{eqnarray}
&& x(N\Delta t) = F(N\Delta t)+(\Delta t)^{1/\alpha}\sum_{k=1}^5 x_k, \label{eq21}\\
&& v(N\Delta t) = F_v(N\Delta t)+(\Delta t)^{1/\alpha}\sum^{5}_{k=1} v_k, \label{eq22}
\end{eqnarray}
where
\begin{eqnarray}
x_k=\sum^{t_{k}/\Delta t}_{i=t_{k-1}/\Delta t}G(i\Delta t)\zeta_{\alpha,1}(N-i), \label{eq23}\\
v_k=\sum^{t_{k+1}/\Delta t}_{i=t_{k-1}/\Delta t}G_v(i\Delta t)\zeta_{\alpha,1}(N-i), \label{eq24}
\end{eqnarray}
where $x_5$ and $v_5$ include all summations for $t > t_4$. In Fig. \ref{fig7} (a)-(d) we show $(\tilde{x}_i, \tilde{v}_i)$ for $i = 1,\cdots,4$ in the phase space. It is interesting to find that for $(\tilde{x}_1, \tilde{v}_1)$ and $(\tilde{x}_3, \tilde{v}_3)$ the distribution of the sample points in the phase space has a preferred direction along with the first and third quadrants, while for $(\tilde{x}_2, \tilde{v}_2)$ and $(\tilde{x}_4, \tilde{v}_4)$ the preferred direction is dominantly in the second and fourth quadrants. This result is easy to understand since in the time domains of $(\tilde{x}_1, \tilde{v}_1)$ and $(\tilde{x}_3, \tilde{v}_3)$ $G(t)$ and $G_v(t)$ have the same sign and in the time domains of $(\tilde{x}_2, \tilde{v}_2)$ and $(\tilde{x}_4, \tilde{v}_4)$ $G(t)$ and $G_v(t)$ have the opposite sign. When all sample points are added according to Eqs. (\ref{eq21}) and (\ref{eq22}), one obtains Fig. \ref{fig7}(e), which shows a homogeneous distribution of the sample points in the phase space of the damped oscillator under the Gaussian noise. After understanding the homogeneous phase space distribution in the underdamped case, we turn into the overdamped case with $\omega = 1, \gamma = 4$. The time-evolution of $G(t)$ and $G_v(t)$ is presented in Fig. \ref{fig1}(c), in which there is no oscillating. It is noted that only one time moment at which $G_v(t)$ changes sign. In this case, strongly anisotropic distributions have been observed in Fig. \ref{fig8}(a) and (b). However, after merging, one again observes an isotropic distribution in the phase space of the damped oscillator in the overdamped case under the Gaussian noise. As a consequence, despite of the underdamped or the overdamped cases, the phase space of the damped oscillator is homogeneous if the noise is Gaussian-like. However, if the noise is L\'evy-like, the situation is quite different, as shown below.

In Fig. \ref{fig9} we consider the phase space distribution of the sample points in the underdamped case under the L\'evy white noise. In the different time intervals, the preferred directions are similar to those under the Gaussian noise, but the difference is that far away from $(\tilde{x},\tilde{v}) = (0, 0)$ there are many scattered points to occur, which is due to the heavy tail distribution of the L\'evy process. When the sample points in different time intervals are merged, the distribution of $(\tilde{x},\tilde{v})$ is almost isotropic, as observed in Fig. \ref{fig9}(e). Likewise, some scattered points are due to the L\'evy noise. In this case, the coordinate and the velocity show weak dependence, which is in consistent with the previous work \cite{Sokolov2011}. However, the distribution is strongly anisotropic in the overdamped case, as shown in Fig. \ref{fig10}(a) and (b). As a consequence, one sees that the distribution of the coordinate and the velocity of the overdamped oscillator is very inhomogeneous under the L\'evy noise, which is presented in Fig. \ref{fig10}(c) and also in Fig. \ref{fig5}(d). In the language of Sokolov et al. \cite{Sokolov2011}, the coordinate and the velocity show strong dependence. According to our analysis, the strong (anti)association originates from the interplay between the damping and the L\'evy noise.

\section{Summary}\label{sec4}

The damped oscillator under the symmetric L\'evy white noise shows inhomogeneous phase space distribution as recovered by Sokolov et al. \cite{Sokolov2011} and it was attributed to the dependence between the coordinate and the velocity under the L\'evy noise. In the present work we explore this dependence and find that it can be explained by considering the interplay between the damping motion of the oscillator and the L\'evy noise. While the underdamped oscillator shows weak anisotropic distribution in the phase space, the overdamped oscillator has a strong inhomogeneous feature, which is in good agreement with the previous result. The essential physics recovered in the present work shed a novel light on the dependence between the coordinate and the velocity under the L\'evy noise and thus provides a way to understand the physics of the damped oscillator under the different noises.

\begin{acknowledgments}
The work is supported by the Program for NCET, NSF and the Fundamental Research Funds for the Central Universities of China.
\end{acknowledgments}


\begin{thebibliography}{99}

\bibitem{West1987} M. F. Shlesinger, B. J. West, and J. Klafter, Phys. Rev. Lett. \textbf{58}, 1100 (1987).
\bibitem{Viecelli1993} J. Viecelli, Phys. Fluids A \textbf{5}, 2484 (1993).
\bibitem{Shlesinger1994} M. F. Shlesinger, G. M. Zaslavsky, and U. Frisch, \textit{L\'{e}vy Flights and Related Topics in Physics} (Springer, New York, 1994).
\bibitem{Janicki1994} A. Janicki and A. Weron, \textit{Simulation and Chaotic Behaviour of Alpha-Stable Stochastic Processes} (Mercel Dekker, New York, 1994).
\bibitem{Zimbardo1995} G. Zimbardo, P. Veltri, G. Basile, and S. Pricipato, Phys. Plasmas \textbf{2}, 2653 (1995).
\bibitem{Baranger2002} M. Baranger, Physica \textbf{305}, 27 (2002).
\bibitem{Dybiec2004} B. Dybiec and E. Gudowska-Nowak, Phys. Rev. E \textbf{69}, 016105 (2004).
\bibitem{Ditlevsen1999a} P. D. Ditlevsen, Geophys. Res. Lett. \textbf{26}, 1441 (1999).
\bibitem{Ditlevsen1999b} P. D. Ditlevsen, Phys. Rev. E \textbf{60}, 172(1999).
\bibitem{Chechkin2001} A. V. Chechkin, V. Y. Gonchar, J. Klafter, R. Metzler, and L. V. Tanatarov, Chem. Phys, \textbf{284}, 233 (2002).
\bibitem{Chechkin2008} A .V. Chechkin, V. Y. Gonchar and M. Szydlowski, Phys. Plasmas \textbf{9}, 78 (2002).
\bibitem{Chechkin2003} A. V. Chechkin, J. Klafter, V. Y. Gonchar, R. Metzler, and L. V. Tanatarov, Phys. Rev. E \textbf{67}, 010102 (2003).
\bibitem{Yi-Zhong Zhuo2005} Jing-Dong Bao, Hai-Yan Wang, Ying Jia, and Yi-Zhong Zhuo, Phys. Rev. E \textbf{72}, 051105 (2005).
\bibitem{Dybiec2007} B. Dybiec, E. Gudowska-Nowak, and I. M. Sokolov, Phys. Rev. E \textbf{76}, 041122 (2007).
\bibitem{Chechkin2007} A. V. Chechkin, O. Y. Sliusarenko, R. Metzler, and J. Klafter, Phys. Rev. E \textbf{75}, 041101 (2007).
\bibitem{Patel1993} A. Patel and B. Kosko, IEEE Trans. Neural Netw. \textbf{19}, NO. 12 (2008).
\bibitem{Dybiec2010a} B. Dybiec, I. M. Sokolov, and A. V. Chechkin,  J. Stat. Mech. \textbf{P07008} (2010).
\bibitem{Dybiec2010b} B. Dybiec, J. Chem. Phys. \textbf{133}, 244114 (2010).
\bibitem{Olemskoi2010} A. I. Olemskoi, S. S. Borysovb, I. A. Shuda, arXiv. 0910.2018 (2010).
\bibitem{Sokolov2011} I. M. Sokolov, W. Ebeling, B. Dybiec, Phys. Rev. E \textbf{83}, 041118 (2011).
\bibitem{West1982} B. J. West and V. Seshadri, Phys. A \textbf{113}, 203 (1982).
\bibitem{Chandrasekhar1943} S. Chandrasekhar, Rev. Mod. Phys. \textbf{15}, 1 (1943).
\bibitem{Sliusarenko2007} A. Yu. Sliusarenko and A. V. Chechkin, Probl Atom Sci Tech. \textbf{N3 (2)}, 293 (2007).
\bibitem{Chambers1976} J. M. Chambers, C. L. Mallows and B. W. Stuck, J. Amer. Statist. Assoc. \textbf{71}, 340 (1976).
\bibitem{Leccardi2000} M. Leccardi, \textit{Comparison of three algorithms for L\'{e}vy noise generation, in: ENOC'05} (Fifth EUROMECH Nonlinear Dynamics
Conference, Mini Symposium on Fractional Derivatives and Their Applications, 2005).
\bibitem{Dybiec2006} B. Dybiec , E. Gudowska-Nowak and P. H\"{a}nggi, Phys. Rev. E \textbf{73}, 046104 (2006).
\bibitem{Janicki1996} A. Janicki, \textit{Numerical and Statistical Approximation of Stochastic Differential Equations with Non-Gaussian Measures} (Wroc{\l}aw: Hugo Steinhaus Centre for Stochastic Methods).

\end{thebibliography}
\end{document}